\documentclass[pra,twocolumn,showpacs,floatfix]{revtex4}
\usepackage{graphicx}
\usepackage{color}
\usepackage{amsmath}
\begin{document}

\title{Rotating Bose-Einstein condensates: Closing the gap between exact and mean-field 
solutions}

\author{J. C. Cremon$^1$, A. D. Jackson$^2$, E. \" O. Karabulut$^{1,3}$, G. M. Kavoulakis$^4$, 
B. R. Mottelson$^2$ and S. M. Reimann$^1$}
\affiliation{$^1$Mathematical Physics, Lund Institute of Technology, P. O. Box 118, 
SE-22100 Lund, Sweden \\
$^2$The Niels Bohr Institute, and Niels Bohr International Academy, Blegdamsvej 17, 
Copenhagen \O, Denmark \\
$^3$Department of Physics, Faculty of Science, Selcuk University, TR-42075 Konya, Turkey\\
$^4$Technological Education Institute of Crete, P. O. Box 1939, GR-71004, Heraklion,
Greece}
\date{\today}

\begin{abstract}

When a Bose-Einstein condensed cloud of atoms is given some angular momentum, it forms vortices 
arranged in structures with a discrete rotational symmetry. For these vortex states, the Hilbert 
space of the exact solution separates into a ``primary" space related to the mean-field 
Gross-Pitaevskii solution and a ``complementary" space including the corrections beyond mean-field. 
Considering a weakly-interacting Bose-Einstein condensate of harmonically-trapped atoms, we 
demonstrate how this separation can be used to close the conceptual gap between exact solutions for  
systems with only a few atoms and the thermodynamic limit for which the mean-field is the correct 
leading-order approximation. Although we illustrate this approach for the case of weak interactions, 
it is expected to be more generally valid.
\end{abstract}


\maketitle

\section{Introduction}
Cold atomic quantum gases are typically dilute with an average interatomic distance much larger 
than the scattering length for atom-atom elastic collisions. This justifies the use of the 
mean-field approximation, which assumes a simple product form for the many-body wavefunction in 
the case of bosonic atoms. The complicated many-body problem is then reduced to one of a single 
variable, and the effect of interactions is described by a nonlinear term. This procedure was 
developed by Gross and Pitaevskii some decades ago \cite{gross1961,pitaevskii1961}. Since the 
experimental realization of Bose-Einstein condensation (BEC) in trapped atomic gases, this 
approach has been used with remarkable success, see e.g., Refs.\,\cite{dalfovo1999, leggett2001, 
pethicksmith2002, pitaevskiistringari2003}. 

One of the many fascinating effects associated with the superfluid properties of these gases 
is the formation of vortices in response to rotation. When the ratio of angular momentum to 
particle number increases, the number of vortices in the cloud grows, and they group in structures 
with discrete rotational symmetries (as illustrated by the mean-field densities in Fig.\,\ref{fig:1} 
below). Such vortex states have been observed in a number of experiments, see for example, 
Refs.\,\cite{chevy2000, madison2000, madison2001, haljan2001, aboshaer2001, engels2002, engels2003, 
schweikhard2004}. The literature on this topic is extensive, as summarized by the reviews 
\cite{bloch2008, viefers2008,cooper2008,fetter2009, saarikoski2010}. For a dilute and 
harmonically-trapped Bose-Einstein condensate of atoms, the {rotational properties have been} 
thoroughly analyzed both within the Gross-Pitaevskii approximation as in Refs.\,\cite{butts1999, 
linn1999, kavoulakis2000, linn2001, garcia2001, vorov2005}, and beyond the mean-field approximation 
as in Refs.\,\cite{wilkin1998, mottelson1999, bertsch1999, kavoulakis2000, jackson2000, smith2000, 
papenbrock2001, huang2001, jackson2001, liu2001, manninen2005, reimann2006a, reimann2006b, cooper2008, 
parke2008, romanovsky2008, liu2009, dagnino2009a, dagnino2009b,papenbrock2012,cremon2013}. 

Going beyond the mean-field approximation one often applies  the so-called configuration-interaction 
(CI) formalism. In this numerical approach, one typically uses the Fock states constructed from a 
given set of single-particle states as a basis for the expansion of the  exact many-body wavefunction.  
Other approaches are often variational, such as quantum Monte-Carlo \,\cite{Pollet2012} or density 
functional techniques for correlated Bose gases\,\cite{Malet2014}. The so-called coupled-cluster 
approach, originally formulated for nuclei \cite{coester1958, coester1960} and often applied to 
atomic and molecular systems of fermions \cite{bartlett1989, bishop2003} has also been adapted to 
bosonic systems \cite{cederbaum2006} and is based on a series expansion of excitation operators 
acting on the corresponding mean-field ground state configuration. 

An important advantage of the full CI approach is that apart from an almost always inevitable 
truncation of the Hilbert space, no further assumptions are made regarding the functional 
form of the many-body wavefunction. The method fully accounts for the correlations between 
the particles and accurately describes the low-lying excitations. However, the dimension of the 
Hamiltonian matrix grows very rapidly with the number of particles. Thus, little is currently 
known about the intermediate regime between small and large systems, in which exact diagonalization 
becomes prohibitively difficult but the mean-field approach still suffers from significant finite-size 
corrections. 

Here, we wish to shed new light on this problem, suggesting a procedure that offers direct insight 
into the question of how a finite-size system of bosonic particles approaches the thermodynamic limit 
in which the Gross-Pitaevskii approach is known to be exact \cite{lieb2000, jackson2001, lieb2009, 
lewin2009}.  With increasing particle number, we find a power-law convergence of the exact ground 
state into the mean-field Gross-Pitaevskii solution. Our study thus provides a clear and general 
strategy for this problem and offers strong arguments for both its validity and practicality. Further, 
these arguments are not limited by either the particle number or by the strength of the interaction. 

\medskip

\begin{figure}[h]
\includegraphics[width=0.75\columnwidth ]{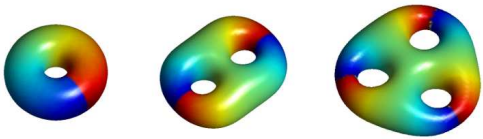}
\caption{Iso-surfaces of the density distribution of the bosonic cloud within the mean-field 
approximation (taken at about one third of the maximum value) for different ratios of angular 
momentum to particle number, $L/N = 1, 1.8$, and 2.3 from left to right (first discussed by 
Butts and Rokshar\,\cite{butts1999}). The vortices appear as holes, and the phase of the order 
parameter (in colorscale from red to blue) jumps by $2 \pi$ when encircling each vortex.}
\label{fig:1}
\end{figure}

\section{Rotational properties within the Gross-Pitaevskii approximation} 
\begin{center}
\begin{table}[b]
\begin{tabular*}{0.45\textwidth }{lccccccc}\hline\hline
$L/N\qquad $&  $c_0$    &  $c_1$   &  $c_2$   &  $c_3$   &  $c_4$   &  $c_6$ \\\hline
\,\,1.0      &    0         &  1.0000 &      0      &     0        &     0      &     0       \\
\,\,1.8      &    0.4803&       0     &   0.7992&     0        &-0.3611 & -0.0127 \\
\,\,2.3      &    0.5312& 0 & 0 & 0.8179 & 0 & -0.2210 \\\hline\hline
\end{tabular*}
\caption{Coefficients $c_m$ of the expansion of the Gross-Pitaevskii order parameter in the lowest 
Landau level for $0 \le m \le 6$ (see text).}
\end{table}
\end{center}
\vskip-0.6truecm
We begin by briefly reviewing some of the well-known results regarding the formation of vortices 
in a weakly-interacting dilute Bose-Einstein condensate. The criterion adopted here for the formation 
of vortices is their energetic stability, namely the minimization of the energy either for a fixed 
value of the angular momentum or for a fixed value of the rotational frequency of the trap. We 
consider a harmonic trapping potential that is very tight along the axis of rotation, here chosen 
to be the $z$ axis, with oscillator frequencies $\omega = \omega _x = \omega _y \ll \omega _z$. The 
two-body potential is assumed to be the usual contact interaction. For sufficiently weak interactions, 
$\hbar \omega _z$ is much larger than the interaction energy, and the atoms occupy only the ground 
state of the harmonic potential in the $z$-direction. Thus, the problem becomes effectively 
two-dimensional. The interatomic potential has the form $V({\bf r}_i - {\bf r}_j) = u_0 \, 
\delta({\bf r}_i - {\bf r}_j)$, with $u_0 = U_0 \int |\phi(z)|^4 \, dz$. Here $U_0 = 4 \pi \hbar^2 
a/M$ is the matrix element for elastic collisions, $a$ is the corresponding scattering length, $M$ 
is the mass of the particles, and $\phi(z)$ is the ground state of the oscillator potential along the 
$z$ axis. The Hamiltonian that we consider is thus
\begin{eqnarray}
 {\hat H} = \sum_{i=1}^N - \frac {\hbar^2 \nabla_i^2} {2 M}
 + \frac M 2 \omega^2 (x_i^2 + y_i^2) + \frac {u_0} 2 \sum_{i \neq j=1}^N 
 \delta({\bf r}_i - {\bf r}_j).
\nonumber \\ 
\label{Hamiltonian}
\end{eqnarray}
 
Within the mean-field Gross-Pitaevskii approximation the many-body wavefunction is assumed
to have a product form, while the corresponding order parameter $\Psi_{\rm MF}(x,y)$ can be 
expanded in the eigenstates of the non-interacting problem. Making the assumption that the 
interaction energy is much smaller than $\hbar \omega $, it is sufficient to consider only the 
single-particle states $\phi _{0,m}$ of the lowest Landau level with zero radial nodes and 
angular momentum $m \hbar \ge 0$. (As we will explain in Sec.\,IV, the assumption of weak 
interactions is not essential, and the approach presented below is expected to remain valid 
for stronger interactions). The order parameter $\Psi_{\rm MF}$ is thus expanded in the basis 
of the states $\phi _{0,m}$, $\Psi_{\rm MF} = \sum_{m \ge 0} c_m \phi_{0,m}$, where the 
amplitudes $c_m$ are variational parameters. 

Minimizing the energy functional subject to the constraint of fixed $L/N$ (where $L \hbar$ is 
the total angular momentum, and $N$ the number of particles in the trap), one finds that as 
$L/N$ increases, there is a sequence of phase transitions associated with the formation of one 
or more vortices in the gas (see the work by Butts and Rokshar \cite{butts1999} and, for example, 
Refs.\,\cite{linn1999, kavoulakis2000, linn2001, garcia2001, dagnino2009a}). This is a direct 
consequence of the fact that for a fixed value of $L/N$ only certain single-particle states are 
occupied by a macroscopic number of atoms of ${\cal O}(N)$. For example, for $L/N = 1$ the 
mean-field approximation yields a solution in which the only occupied single-particle state is 
the one with $m=1$ and there is a single vortex at the trap center (see Refs.\,\cite{{wilkin1998, 
mottelson1999, butts1999, bertsch1999, kavoulakis2000}}). A similar behavior is found at higher 
values of $L/N$ \cite{butts1999, kavoulakis2000}. For example, for $L/N = 1.8$, only the 
single-particle states with $m = 0, 2, 4, \dots$ are occupied and the order parameter has two-fold 
symmetry. Correspondingly, for $L/N = 2.3$, the mean-field state consists of single-particle 
orbitals with only $m = 0, 3, 6, \dots$ and it has three-fold symmetry.\,\,\footnote{We note that 
generally there is a saturation in the necessary size of the basis with increasing single-particle 
angular momentum $m$. This is shown in Fig.\,5 for the example of the two-vortex state, where the 
occupancy of orbitals with $m>4$ is reduced by orders of magnitude.}

The actual values of the variational parameters $c_m$ which are derived within this method are given 
in the Table below for $L/N= 1.0, 1.8$ and $2.3$ (with $0 \le m \le 6$). In Fig.\,\ref{fig:1} we show 
the density isosurfaces for these states (having one-fold, two-fold, and three-fold symmetry), where 
the color scale (from red to blue) indicates that the phase of the order parameter changes by $2\pi$ 
when encircling the vortex singularity \cite{butts1999}. 

\section{Exact solutions compared to mean-field} 

In order to obtain the exact solution of the problem\,\cite{wilkin1998, bertsch1999, jackson2001}, 
one diagonalizes the many-body Hamiltonian ${\hat H}$ subject to the constraints of fixed particle 
number, $\sum_m n_m = N$, and fixed total angular momentum, $\sum_m m n_m = L$. With increasing 
values of $N$ and $L$, the exponentially growing computational complexity severely restricts the 
size of numerically tractable systems to a few dozen atoms at most. Yet, studying the detailed 
structure of the exact wavefunction suggests a substantial simplification: In analogy with the 
mean-field approach discussed above, when an yrast state (i.e., the state with lowest energy at 
fixed angular momentum $L$) has a given discrete rotational symmetry, only certain single-particle 
states are occupied by a macroscopic number of atoms of ${\cal O}(N)$.  This permits a separation 
of the total Hilbert space into a ``primary" subspace that includes only Fock states constructed 
exclusively from single-particle states which are macroscopically occupied within the mean-field 
approximation and a far larger ``complementary" space that consists of all other Fock states involving 
single-particle orbitals outside the mean-field space \cite{cremon2013}. The inclusion of this 
complementary space leads to corrections to the mean-field energy that are of higher order in 
$1/N$ relative to the contribution from the primary space \cite{jackson2001}. We shall show in the 
following that this fact can be exploited efficiently to bridge the gap between the few-body and 
thermodynamic limits.   

\subsection{The ``unit vortex"} 

We begin with the relatively simple case of $L/N = 1$, where (as described above) within 
the mean-field approximation only the $m=1$ orbital is occupied. As a result there is a single 
vortex located at the centre of the trap. The primary space thus consists of only one Fock 
state, with all atoms occupying the $m=1$ orbital. The complementary space is spanned by all other 
Fock states that have a non-zero occupancy of any single-particle state with $m \neq 1$. If one 
works within a truncated space including only the orbitals with $m = 0, 1$ and $2$, the yrast state 
$|\Psi_0 \rangle $ is known analytically to leading order in $N$ \cite{wilkin1998, kavoulakis2000},  
\begin{eqnarray}
 | \Psi_0 \rangle = \sum_k \frac {(-1)^k} {\sqrt {2}^{k+1}} |0^k, 1^{N-2k}, 2^k \rangle, 
\label{fock1}
\end{eqnarray}
where the ket on the right denotes the Fock state with single-particle states in the lowest Landau 
level with $m=0, 1$ and 2, and corresponding occupancies noted by the exponents. Returning to
the separation of the full Hilbert space which we described above, the primary space consists of the 
single state with $k=0$, with a probability $1/2$, while all the other states with $k \neq 0$ 
constitute the complementary space, with a probability also equal to 1/2. Note that the amplitudes 
in Eq.\,(\ref{fock1}) decrease exponentially with $k$, i.e., with the occupancy of the states 
belonging to the complementary space. As we will see below, this is a more general feature that 
also appears for larger values of the angular momentum. 

We now consider a more systematic analysis of the convergence of the exact solution as a function 
of a gradual increase of the contribution from the complementary space. In order to capture fully 
the finite-size corrections, we extend the truncated space used in Eq.\,(\ref{fock1}) by including 
the states with $0 \le m \le 6$ which necessitates the use of numerical methods. We evaluate the 
many-body states for a fixed number of particles $n_1$ in the $m = 1$ state (which is the occupation 
of the primary space) with the remaining $(N - n_1)$ particles in the complementary space. A 
Hamiltonian matrix is constructed for each value of $(N-n_1)$, and the single eigenstate of lowest 
interaction energy is selected from each matrix. A truncated Hamiltonian is built and diagonalized 
in this new basis of lowest-energy states to obtain the approximate energy spectrum $E_{\rm A}^i$ 
and the corresponding eigenfunctions $|\Psi_{\rm A}^i \rangle$ (where the index $i = 0, 1, 2, \dots$ 
labels the excited states), here for the example of $N = 100$ atoms. 

Figure 2 shows the low-energy spectrum as evaluated within this scheme, as a function of 
the highest occupancy of the complementary space $N_c$ (red circles). (The energy is given in 
units of $v_0=u_0\int \mid\!\!\phi _{00}\!\!\mid ^4d^2 r$, where $\phi _{00}$ is the single-particle 
ground state of the two-dimensional harmonic oscillator). The right side of this graph also shows 
the energy spectrum evaluated within the usual full diagonalization of the many-body Hamiltonian 
(blue circles) with the same truncation, $0 \le m \le 6$. 

We see that there is a rapid convergence of the approximate solution to the exact result. Low-lying 
excited states are also reproduced fairly well, with larger deviations being apparent only in the 
higher-energy section of the spectrum. The relative error between the eigenenergies as evaluated 
within our model and with full diagonalization (in the same subspace) decreases exponentially with 
$N_c$. {Remarkably, our method reproduces the yrast state as well as the low-lying excited states 
with high accuracy.} As shown in panel (a) of Fig.\,3 for the yrast state (``Y") and the {first 
non-trivial excited state ``G" that is related to the Goldstone mode} \cite{ueda2006}, relative 
errors as small as $10^{-7}$ to $10^{-10}$ are obtained for values of $N_c \approx N/2$.  (The 
first excited state, labeled ``CM", is a {trivial} center-of-mass excitation). Panel (b) of Fig.\,3 
shows a logarithmic plot of the deviation from unity overlap of the model yrast state and the exact 
yrast state, $1-\mid \!\langle \Psi _{\rm A}^0\mid \!\Psi _{\rm ex}\rangle \mid^2$, as a function 
of $N_c$. This plot clearly shows an exponential convergence with the number of particles in the 
complementary space. Finally, Fig.\,3 (c) shows the size of the submatrices arising in our 
calculations.  Note that all of these submatrices are dramatically smaller than that of the full 
CI matrix, which has a dimension of $189,509$ for the specific example considered here.  

\begin{figure}[t]
\includegraphics[width=0.9\columnwidth]{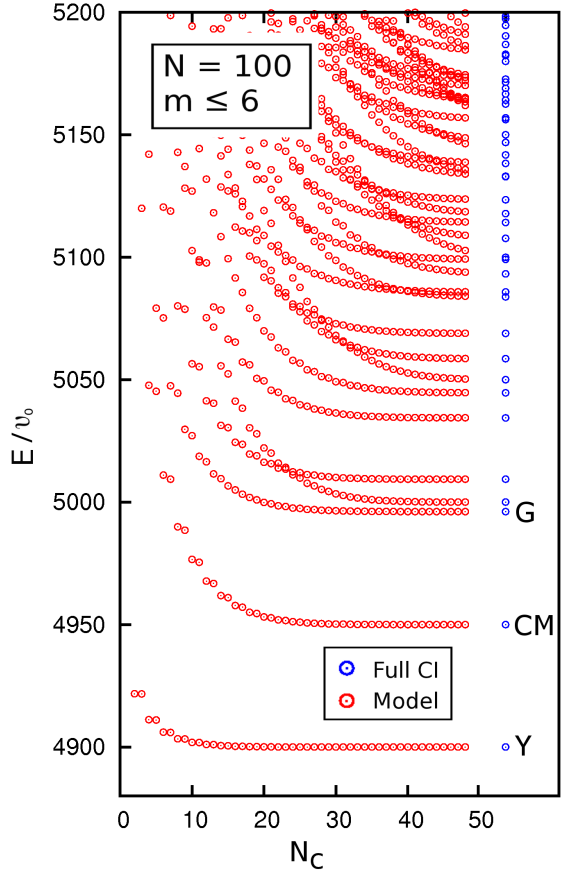}
\caption{Interaction energies (in units of $v_0$, see text) for $L = N = 100$ evaluated within the 
model (left, red circles) as a function of the highest occupancy $N_c$ of the complementary space 
for a single-particle basis of orbitals with $0 \le m \le 6$. The blue bullets to the right show 
the result from the ``exact" approach in the same subspace of single-particle states (see text). 
The relative error between the model and the full diagonalization in the same subspace is given 
in panel (b) of Fig.\,3 for the yrast state (``Y") and the Goldstone mode (``G") {(see text)}, 
converging to numbers as small as $10^{-7}$ to $10^{-10}$ for values of $N_c \approx N/2$. (The 
state labeled "CM" is a center-of-mass excitation). }
\label{fig2}
\end{figure}

\begin{figure}[t]
\begin{center}
\includegraphics[width=0.7\columnwidth]{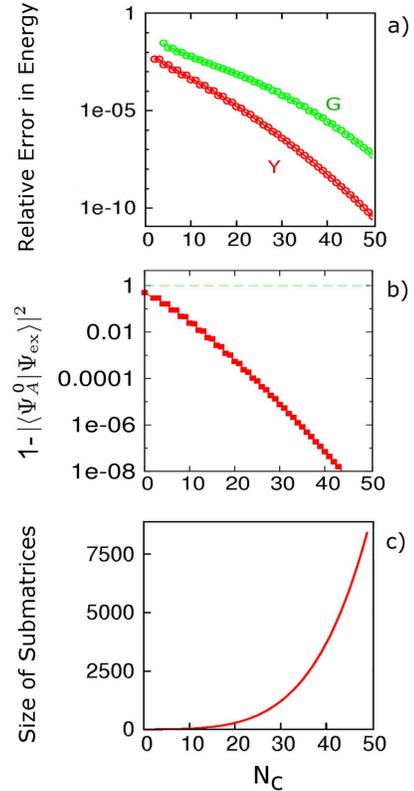}
\end{center}
\caption{(a) Relative error in the interaction energy, and (b) the deviation from unity overlap 
of the model yrast state with the exact yrast state, $1-\mid \!\langle \Psi _{\rm A}^0\mid 
\!\Psi _{\rm ex}\rangle \mid^2$, on a logarithmic scale as a function of the highest occupancy 
of the complementary space, $N_c$, for the data of Fig.\,2.  Panel (c) shows the dimension of the 
submatrices as a function of $N_c$.} 
\label{fig3}
\end{figure}

\subsection{States of discrete rotational symmetry} 

While the above example of the unit vortex at $L/N=1$ is instructive, it is special in the sense 
that within the mean-field approximation there is only {\it one} state that is macroscopically 
occupied.  We therefore consider, without loss of generality, the ratio $L/N=9/5$ where within 
the mean-field approximation {\it two} vortices have nucleated in the cloud. 

In what follows below we show how our method is applied considering this case as an example.
We start with some approximate and semi-analytic results, which demonstrate the use of our method.
At the end of this section we present the full numerical results. 

In order to simplify the discussion, we first truncate the space to the single-particle states with 
$0 \le m \le 4$. (It is straightforward to generalize the arguments presented below to larger spaces). 
For large $N$, there is a macroscopic occupancy of the single-particle states with $m = 0,\ 2$ and 
$4$ in the primary space. In this limit, it is convenient to approximate the exact yrast state by 
the sum 
\begin{eqnarray}
  |\Psi_0 \rangle &=& \sum_k (-1)^k \psi_k |0^{n_0(k)}, 2^{n_2(k)}, 4^{n_4(k)} \rangle,
\label{yrast}
\end{eqnarray}
where the occupancy of the orbitals with $m = 0, 2$ and $4$ are $n_0(k)$, $n_2(k)$ and $n_4(k)$, 
respectively. Fixing the occupancy of one orbital, for example, $n_0(k) = k$, the constraints 
of fixed particle number and fixed total angular momentum determine the occupancy of the other 
two. Consequently, for the above choice of $n_0(k)$, we have $n_2(k) = 11 N/10 - 2 k$ and 
$n_4(k) = k - N/10$. As discussed in Ref.\,\cite{jackson2001}, within the states contained 
in Eq.\,(\ref{yrast}) the eigenvalue equation takes the form 
\begin{equation}
  - V_{k,k-1} \psi_{k-1} + V_{k,k} \psi_k - V_{k,k+1} \psi_{k+1} = E \psi_k.
\end{equation}
Here $V_{k,k'}$ are the matrix elements of the interaction $V$ between the states $|k \rangle$ and 
$|k' \rangle$ on the right of Eq.\,(\ref{yrast}), and $E$ is the interaction energy. The above
equation is obvious, since in addition to the diagonal matrix elements, the interaction, being a
two-body operator, connects states where two atoms from the states with $m = 0$ and $m = 4$ get
transferred to the state with $m = 2$, and vice versa.

Assuming that $\psi_k$ is a smooth and differentiable function of $k$, this eigenvalue equation 
assumes the familiar form  
\begin{eqnarray}
   - \frac 1 2 (V_{k,k-1} + V_{k,k+1}) \partial_k^2 \psi_k  +  V_{\rm eff}(k) \psi_k = E \psi_k,
\end{eqnarray}
{where $V_{\rm eff}(k) = V_{k,k}- V_{k,k-1} - V_{k,k+1}$.  In the vicinity of its minimum at 
some $k_0$ where $V_{\rm eff}$ has the value $E_0$, this effective potential can be approximated 
as $V_{\rm eff} - E_0 \propto (k - k_0)^2/2$. }  Thus, $\psi_k$ satisfies an eigenvalue equation 
of a linear single-particle problem in the effective potential $V_{\rm eff}(k)$. For the example of 
$L/N = 9/5$, in the limit of very large $N$ we find that the minimum of $V_{\rm eff}(k)$ occurs 
at $k_0/N = n_0/N \approx 0.2289$, a number that is very close to the value of the mean-field 
coefficient for $|c_0|^2 (\approx 0.2307$) given in the Table above (albeit here for $0 \le m 
\le 4$). The energy of the yrast state within the primary space is found to be $E \approx (0.1880 
N^2 - 0.3149 N) v_0$ plus terms of order unity. This expression for $E$ must now be corrected at 
${\cal O}(N)$ for the effects of the complementary space.

To gain insight into the role of the orbitals outside the mean-field space, we  first consider 
contributions to the complementary space due to the $m=1$ single-particle state only. The $\psi_k$ 
in Eq.\,(\ref{yrast}) for the yrast state is a Gaussian with a width of ${\cal O}(\sqrt N)$. If one 
promotes $2 n_1$ particles to the single-particle state with $m=1$, where $n_1$ is of ${\cal O}(N^0)$, 
to leading order in $N$, the corresponding yrast state can be written as
\begin{eqnarray}
  |2 n_1 \rangle \propto (a_1^{\dagger} a_1^{\dagger} 
  a_0 a_2)^{n_1} |\Psi_0 \rangle.
\label{yrastc}
\end{eqnarray}
Here $|2 n_1 \rangle $ denotes the yrast state with $2n_1$ atoms in the single-particle state 
with $m=1$  and $a_m^{\dagger}$ and $a_m$ are the usual creation and annihilation operators for 
a particle with angular momentum $m \hbar$. The Gaussian form of the amplitudes of the primary 
space components of the state $|2 n_1\rangle$ is preserved, and its center is simply shifted by 
a term of order unity. As before, the primary components are all of ${\cal O}(N)$ with a width 
of ${\cal O}(\sqrt{N})$, and the occupancy of the states in the complementary space is of 
${\cal O}(N^0)$. As a consequence, the energy of the state $|2 n_1\rangle$ is the same as that 
of $|\Psi_0 \rangle$ to leading order in $N$, i.e., to ${\cal O}(N^2)$, and it is only necessary 
to consider corrections from the complementary space which are of subleading order, i.e., 
${\cal O}(N)$. This implies that it is sufficient to approximate the full state of 
Eq.\,(\ref{yrast}) by the single component $|\Psi_0 \rangle = |0^{n_0(k_0)}, 2^{n_2(k_0)}, 
4^{n_4(k_0)} \rangle$. This is a very considerable simplification. 

Using this single component (appropriately renormalized to unity), we find that neither the 
curvature of $V_{\rm eff}$ nor the ``inertial parameter'' $(V_{k,k-1} + V_{k,k+1})$ depend on 
$n_1$ to leading order. Further, we find that the diagonal energies scale linearly with $n_1$, 
while the off-diagonal matrix elements $\langle 2n_1, 0 | V | 2 n_1 + 2, 0 \rangle$ are seen to 
be proportional to $N \sqrt{(2 n_1 + 1)(2 n_1 +2)}$. These are the only nonzero off-diagonal matrix 
elements which come from the operator $a_1 a_1 a_0^{\dagger} a_2^{\dagger}$ (plus its Hermitian 
conjugate). Diagonalizing the resulting matrix, we find that the yrast energy becomes $E \approx 
(0.1880 N^2 - 0.3885 N) v_0$ and that the probabilities of the various states with $2 n_1$ atoms 
in the $m=1$ state decrease exponentially with $n_1$. We emphasize here that the above result is 
generic and not specific for the example that we have considered (see Appendix). 

Let us now turn to the case where contributions from both the $m = 1$ and $m=3$ single-particle 
orbitals are included in the complementary space. Generalizing Eq.\,(\ref{yrastc}), we see that 
the states including contributions from the complementary space can be constructed as 
\begin{displaymath}
  |2 n_1+1, 2 n_3+1 \rangle  \propto a_1^{\dagger 2 n_1 + 1} 
  a_3^{\dagger 2 n_3 + 1} a_0^{n_1} a_2^{n_1 + n_3 + 2} a_4^{n_3} 
  |\Psi_0 \rangle,
 \end{displaymath}
 \begin{eqnarray}
  |2 n_1, 2 n_3 \rangle \propto a_1^{\dagger 2 n_1} 
  a_3^{\dagger 2 n_3} a_0^{n_1} a_2^{n_1 + n_3} a_4^{n_3} 
  |\Psi_0 \rangle. 
\label{yrastcc}
\end{eqnarray}
We emphasize that the counting of states in this equation is correct: Naively, one might expect 
that two distinct states, e.g., with $n_1 = n_3 = 1$ could be constructed from $|\Psi_0 \rangle$ 
through the application of either $a_1^{\dagger} a_3^{\dagger} a_2 a_2$, or $a_1^{\dagger} 
a_3^{\dagger} a_0 a_4$. Given the Gaussian nature of $\psi_k$, however, these states are not 
orthogonal, and only one of them should be included.  

There are four classes of nonzero off-diagonal matrix elements. The first two classes include
\begin{eqnarray}
\langle n_1+2, n_3 |\,V\ | n_1, n_3 \rangle = \sqrt{n_0\,n_2}
  V_{11,20} \sqrt{(n_1+1)(n_1+2)},
  \nonumber 
\end{eqnarray}
and
\begin{eqnarray}
\langle n_1, n_3 + 2 | V | n_1, n_3 \rangle = \sqrt{n_2\, n_4}
  V_{33,24} \sqrt{(n_3 + 1)(n_3 + 2)},
\nonumber 
\end{eqnarray}
where $V_{ij,kl} = \langle \phi_{0i}, \phi_{0j} | V | \phi_{0k}, \phi_{0l} \rangle$. The remaining 
two classes reflect the staggering of the ground state wave function (i.e., the alternating sign seen 
explicitly in Eq.\,\ref{fock1} above). They have the form
\begin{eqnarray}
   \langle n_1, n_3 | V | n_1 - 1, n_3 + 1 \rangle = (-1)^{n_3}
\sqrt{n_1 (n_3 + 1)} 
\nonumber \\
\times (2 \sqrt{n_0\, n_2} V_{12,30} - 2 \sqrt{n_2\, n_4} V_{14,30})
\end{eqnarray}
and
\begin{eqnarray}
    \langle n_1 + 1, n_3 + 1| V | n_1, n_3 \rangle = (-1)^{n_3}
\sqrt{(n_1 + 1)(n_3 + 1)} 
\nonumber \\
    \times (n_2 V_{22,13} - 2 \sqrt{n_0\, n_4} V_{04,13}).\ \ \ 
\end{eqnarray}
For $N_c$ particles in the complementary space, the matrix has dimension $N_c/2+1$. Considering 
as an example $N_c = 18$, the yrast energy $E \approx (0.1880 N^2 - 0.3149 N) v_0$ obtained above 
for the primary space is shifted by the amount $\Delta E = -0.1893 N v_0$. As previously discussed for 
the simpler case with only the $m=1$ orbital from the complementary space, the amplitudes show an exponential decrease with $n_1$ and $n_3$.

\begin{figure}[t]
\includegraphics[width=0.8\columnwidth ]{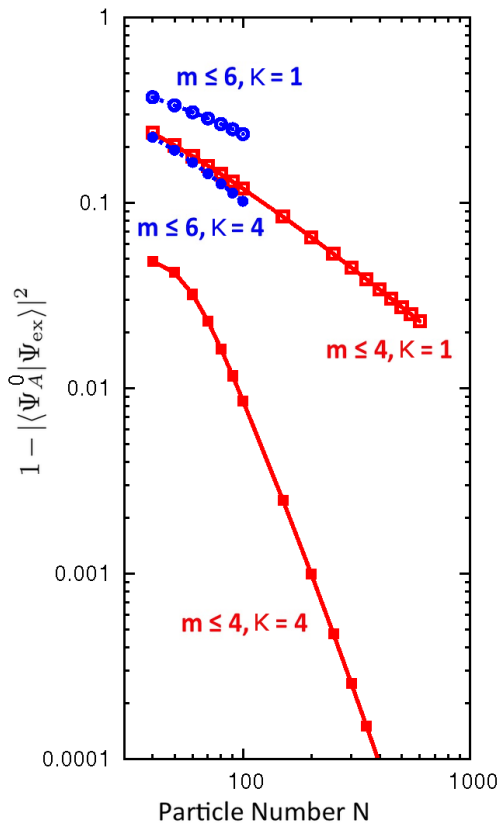}
\caption{The quantity $1 - |\langle \Psi_{A}^0|\Psi_{\rm ex} \rangle|^2$ as a function of $N$
for a fixed angular momentum per particle, $L/N = 9/5$.  Here, at each point the number $N_c$ of 
particles in the complementary space was increased until numerical convergence of the result was 
obtained. The red squares refer to the space  with $0 \le m \le 4$ (open squares, $K=1$, and solid 
squares, $K = 4$, where $K$ is the number of lowest-energy states included in the reduced matrix), 
and the blue circles to the space with $0 \le m \le 6$ (open circles, $K = 1$, and solid circles, 
$K = 4$).}
\label{fig4}
\end{figure}

\begin{figure}[t]
\includegraphics[width=0.75\columnwidth ]{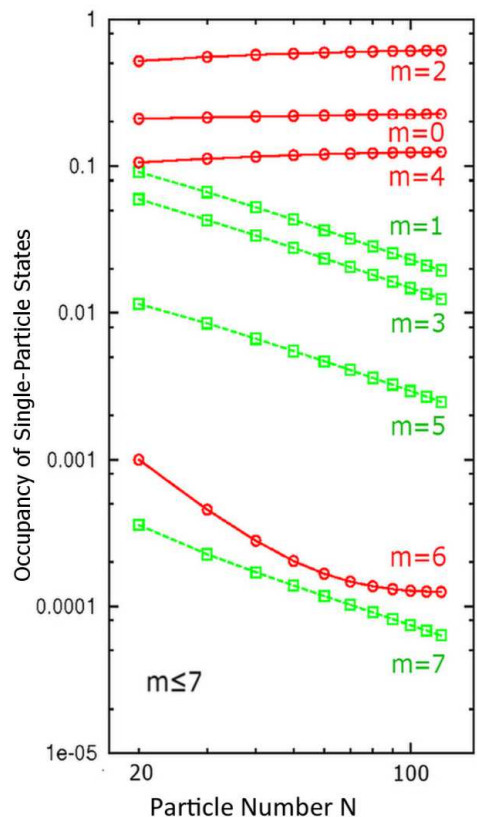}
\caption{Occupancies of the exact yrast states at $L/N=9/5$ as a function of the particle number, 
$N$, on a double-logarithmic scale here for a basis with $0 \le m\le 7$. The saturation of 
the single-particle basis is reflected by the significant reduction in occupancy for higher values 
of $m$. The red circles indicate single-particle states that belong to the primary space; the green 
squares indicate those that belong to the complementary space. Clearly, with increasing $N$ there is 
a convergence towards the occupancies obtained from mean-field (see the Table above).}
\label{fig5}
\end{figure}

In the preceding paragraphs we approximated the yrast state with the single component $|\Psi_0 
\rangle = |0^{n_0(k_0)}, 2^{n_2(k_0)}, 4^{n_4(k_0)} \rangle$ which allowed us to give a 
detailed description of our method, providing us with some semi-analytic results. In what follows 
below we apply the same method fully numerically, without making the approximation for the yrast 
state consisting of a single component and compare it with the result of the full diagonalization 
(in the spirit of the case of $L/N = 1$ discussed above). Here, however, we consider only the 
yrast state, since the analysis of the excitation spectra beyond the value of $L/N = 1$ is more 
complicated due to the fact that, even for the low-lying states, competing solutions of different 
symmetry are found, see Ref.\,\cite{cremon2013}. 

We follow the same procedure as above, considering two different truncations of the Hilbert 
space with $0 \le m \le 4$ and $0 \le m \le 6$. First, we evaluate the many-body states of 
Eq.\,(\ref{yrastcc}) for each configuration of particles in the complementary space. Then, the 
lowest-energy state for each such configuration is chosen, and the truncated Hamiltonian is 
diagonalized in this new basis of lowest-energy states to obtain the approximate ground state 
energy $E_A^0$ and corresponding wavefunction $|\Psi_A^0 \rangle$. Figure 4 shows the quantity 
$1 - |\langle \Psi_{A}^0| \Psi_{\rm ex} \rangle|^2$ versus $N$ in the space $0 \le m \le 4$ 
(squares) and $0 \le m \le 6$ (circles), on a double-logarithmic scale. Intriguingly, for large 
$N$, this quantity shows a simple power-law behaviour, $1 - |\langle \Psi_{A}^0 |\Psi_{\rm ex} 
\rangle|^2 \propto N^{-\gamma}$. 

The fact that the overlap becomes worse for the more extended space, $0 \le m \le 6$, as 
compared to the more restricted one, $0 \le m \le 4$, is not a surprise, since the relative 
dimensionality of the full Hilbert space compared with the dimensionality of the space of our 
approach increases dramatically in the more extended space. The crucial result of this analysis 
is that the Fock state amplitudes for a given configuration in the complementary space decrease 
exponentially with the number of particles in that space in both cases (we also emphasize that 
the number of the single-particle orbitals which are occupied by a macroscopic number of atoms
saturates quickly for larger values of $m$, as it is clearly seen from the logarithmic plot of 
the occupancies in Fig. 5). The exponent is independent of $N$ in the limit $N \to \infty$, 
indicating that the number of particles in the complementary space has a limit of ${\cal O}(N^0)$. 
This behavior is in fact strongly supported by the occupancies obtained through direct 
diagonalization, as seen also in Fig.\,5. The occupancy of the states which construct the 
primary space in compliance with the mean-field approximation (red lines) increases (except 
for the orbital with $m=6$ that saturates), while the orbitals of the complementary states show
a power-law decay with the number of particles.

More generally, one can retain the $K$ eigenstates of lowest energy for each configuration of 
particles in the complementary space. (This corresponds to retaining excited states of the effective 
harmonic-oscillator problem in the primary space).  For a given basis set, an increase of $K$ 
accelerates the convergence towards the full solution (see Fig.\,\ref{fig4}). For fixed $N$, one 
finds exponential convergence in $K$. In this case, however, the exponent depends on $N$ such that 
convergence is more rapid for larger particle numbers, and the generalized oscillator ground state 
{\it alone\/} contributes to the yrast state as $N \to \infty$. We stress that for $N_c = N$ and 
retaining all the possible $K$ states, the approach is simply a passive unitary transformation of 
the basis and the results are necessarily identical to the full exact solution.

\section{Conclusions} 

In short, this paper suggests a significantly simplified understanding of the properties of a 
rotating Bose-Einstein condensate of trapped atoms. The direct numerical strategy for this problem 
would be to include a certain set of single-particle states and to diagonalize the resulting many-body 
Hamiltonian matrix. The difficulty is that the dimension of this matrix grows prohibitively as the 
number of particles or the angular momentum (and thus, consequently, the number of necessary 
single-particle basis states) increases.

The method presented in this study makes use of the fact that only certain single-particle states are 
macroscopically occupied, while all other states have an occupancy of order unity. This introduces a 
natural separation of the Hilbert space into a primary and a complementary part. The first, containing 
the macroscopically occupied single-particle states, can be regarded as a generalised harmonic 
oscillator problem that can withstand major truncation when $N$ is large. The resulting simplification 
is significant: The size of the Hamiltonian matrix can be reduced safely by a factor of order 
$N^{\kappa - 2}$, where $\kappa$ is the number of single-particle states included in the primary 
space. The contribution of the complementary subspace to the many-body states falls exponentially with 
the number of particles in it. Therefore, the present approach shows clearly that the vast majority of 
these states do not make a significant contribution to the yrast states, providing a simple 
understanding of how scattering processes between the primary and complementary spaces govern the 
transition from finite-sizes to the thermodynamic limit. 

At mean-field level there are discontinuous phase transitions between states of different rotational 
symmetry corresponding to level crossings \cite{cremon2013}.  The states involved in such crossings 
can be constructed using the methods described here for distinct choices of the primary space.  In 
the immediate vicinity of the crossing point, these states will be nearly degenerate and can in 
principle mix.  Fortunately, however, the fact that they are based on different primary spaces ensures 
that matrix elements of the interaction between these states will vanish exponentially with the number 
of particles.  Thus, such mixing becomes increasingly unimportant as the number of particles grows 
unless one is precisely at the crossing point.

{The analysis presented here has been restricted to the limit of weak interactions, where one may 
neglect the single-particle eigenstates of the harmonic-oscillator potential with radial nodes. We 
stress, however, that our results are quite general and are {\it not} specific to this perturbative 
regime. Rather, the illustration of our method for the case of weak interactions represents a ``proof
of principle" and provides a representative example of our approach. Even in the regime 
of stronger interactions, which is actually of greater experimental relevance, one can solve the
mean-field Gross-Pitaevskii equation to determine which states should be contained in our primary 
subspace. This subspace could, in principle, contain any or even all Landau levels with an angular
momentum consistent with the discrete rotational symmetry of the ground state. In practice, mean-field 
calculations for stronger interactions show that the probability of finding states with $n_r$ radial
nodes in the mean-field wave function decreases exponentially with increasing $n_r$. The inclusion of 
additional Landau levels will not lead to any material complication in the construction or the 
solution of the generalized harmonic-oscillator problem described in Sec.\,III A. Excited Landau 
levels for other angular momenta will contribute to the complementary space. Since the exponential 
convergence found in the present manuscript is dictated by angular momentum considerations and
not by the radial structure of the single-particle wave functions, this feature will be unaltered 
by the inclusion of higher Landau levels. In short, the scheme introduced here will remain valid and 
useful even if additional Landau levels are included. However, as mentioned above, internal 
convergence criteria must be adopted for assessing the accuracy of such calculations since full 
numerical diagonalizations will not be practical.}

While the truncations adopted here appear to be particularly promising when the number of particles 
is large, we have not proven that the approach is a viable quantitative alternative to the exact 
diagonalization for large systems. Such a claim would require extensive benchmarking that is beyond 
the scope of this work and remains a matter for further investigation. Rather,  the present study 
offers new and very explicit insight into the structure of the many-body wavefunction and its relation 
to the mean-field approximation for a rotating atomic superfluid. Although we  have focused primarily 
on the yrast line, the  procedure adopted here should also be suitable for investigating the richness 
of the excitation spectrum. We stress that the method is physically well-motivated and provides a 
well-defined transformation of the basis of many-body states that is completely passive in the sense 
that it suggests the order but not the degree of truncations of the basis. 

\acknowledgements

We thank C. J. Pethick for discussions and helpful comments. This work was financially supported by 
the Swedish Research Council and the Nanometer Structure Consortium at Lund University, and the POLATOM 
ESF Research network. 

\section*{APPENDIX: Toy Model}

Consider, e.g., a real symmetric matrix which is zero except for the diagonal matrix elements 
$A_{n,n} = n-1$ and the off-diagonal matrix elements $A_{n,n \pm 1}$, with $A_{n,n + 1} = n f$. 
The eigenvalue equation 
\begin{equation}\nonumber
   A_{n,n-1} c_{n-1} + A_{n,n} c_n + A_{n, n+1} c_{n+1} = E c_n
\end{equation}
has the form 
\begin{equation}\nonumber
   (n-1) f c_{n-1} + (n-1) c_n + n f c_{n+1} = E c_n. 
\end{equation}
The lowest eigenvalue of this matrix has a finite value in the limit that its dimension approaches 
infinity, provided that $0 \le |f| \le 1/2$. Subject to this restriction, the lowest eigenvalue is 
$E = -|f|x$ and the corresponding solution is $c_n = x^{n} \sqrt{1-x^2}$ with 
$x = -(1 - \sqrt{1 - 4 f^2})/(2f)$. For large values of $n$, the above eigenvalue equation has 
the simpler form $f c_{n-1} + c_n + f c_{n+1} = 0$, which has the solution $c_n \propto x^n$, 
independent of $E$. In other words, $c_n$ decays exponentially at precisely the same rate for 
all eigenvectors of finite energy. Although elementary, this toy model illustrates the present 
mechanism leading to exponential convergence.

\end{document}